\ifx\mnmacrosloaded\undefined \input mn\fi
\input epsf.tex

\begintopmatter

\title{Kinematically Lopsided Spiral Galaxies}

\author{R.~A.~Swaters, R.~H.~M.~Schoenmakers, R.~Sancisi and T. S. van
  Albada }

\affiliation{Kapteyn Astronomical Institute, P.O. Box 800, 9700 AV
Groningen, The Netherlands}

\shortauthor{R.~A.~Swaters et al.}
\shorttitle{Kinematically Lopsided Spiral Galaxies}

\abstract { Asymmetries in the distribution of light and neutral
  hydrogen are often observed in spiral galaxies.  Here, attention is
  drawn to the presence of large-scale asymmetries in their
  kinematics.  Two examples of kinematically lopsided galaxies are
  presented and discussed.  The shape of the rotation curve --rising
  more steeply on one side of the galaxy than on the other-- is the
  signature of the kinematic lopsidedness.  It is shown that
  kinematic lopsidedness may be related to lopsidedness in the
  potential, and that even a mild perturbation in the latter can
  produce significant kinematical effects.  Probably at least half of
  all spiral galaxies are lopsided. }


\maketitle

\section{Introduction}

Large scale asymmetries in the optical appearance of spiral galaxies
have been known for a long time (e.g., M~101, Arp 1966), but only a few
systematic studies have been carried out.  Baldwin, Lynden-Bell \&
Sancisi (1980) drew attention to lopsided HI distributions of disc
galaxies, emphasizing that the asymmetry affects large parts of the
disc and that it is a common phenomenon among spiral galaxies.  Also,
they proposed a model in which the lopsidedness in a galaxy starts as
a series of initially aligned elliptical orbits in an axisymmetric
potential. Due to differential rotation, the pattern will wind up,
however, and the lopsidedness will slowly disappear.  They estimated
that the typical lifetime of the lopsidedness is between 1 and 5 Gyr.

Richter \& Sancisi (1994) made an estimate of the frequency of
asymmetries from the shape of the HI line profiles. From an inspection
of about 1700 global profiles they found that at least 50\% of the
galaxies have strong or mild asymmetries. This result has recently
been confirmed by Haynes et al.\ (1998) who have obtained new, high
precision global HI profiles for a sample of isolated spirals.

As both the disc kinematics and the HI density distribution determine
the shape of the global profile, it is difficult to assess the nature
of the asymmetry from the global profile alone.  Richter \& Sancisi
(1994) emphasize, however, that the HI distributions and kinematics of
individual galaxies indicate that the profile asymmetries often
originate from a large-scale, structural lopsidedness affecting the
whole disc, confirming the earlier findings of Baldwin et al.\ (1980).

\beginfigure*{1}
\epsfxsize=0.75\hsize
\null\vskip-8.0cm
\null\hskip-0.2cm\epsfbox{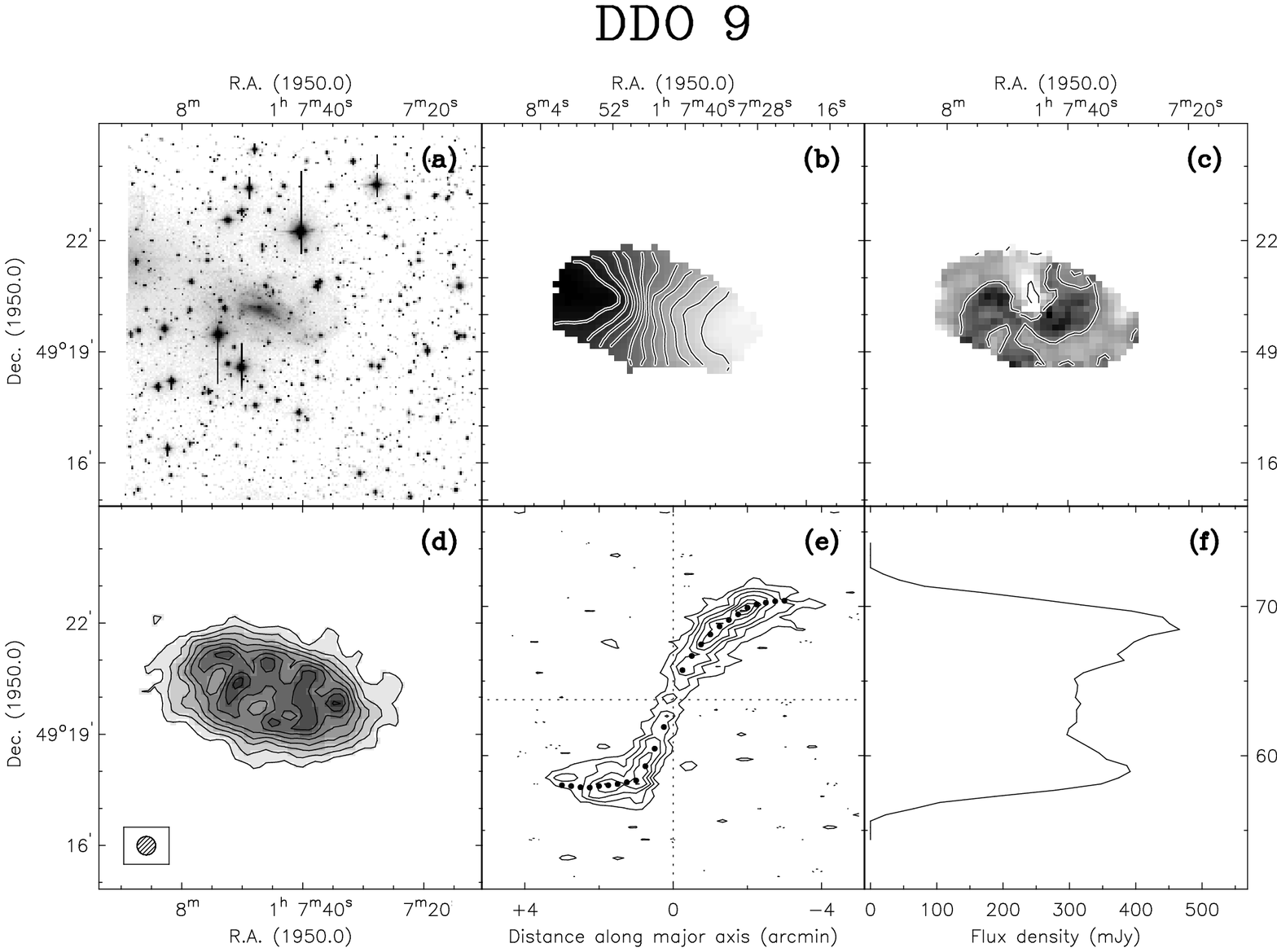}
\epsfxsize=0.75\hsize
\null\vskip-6.0cm
\null\hskip-0.2cm\epsfbox{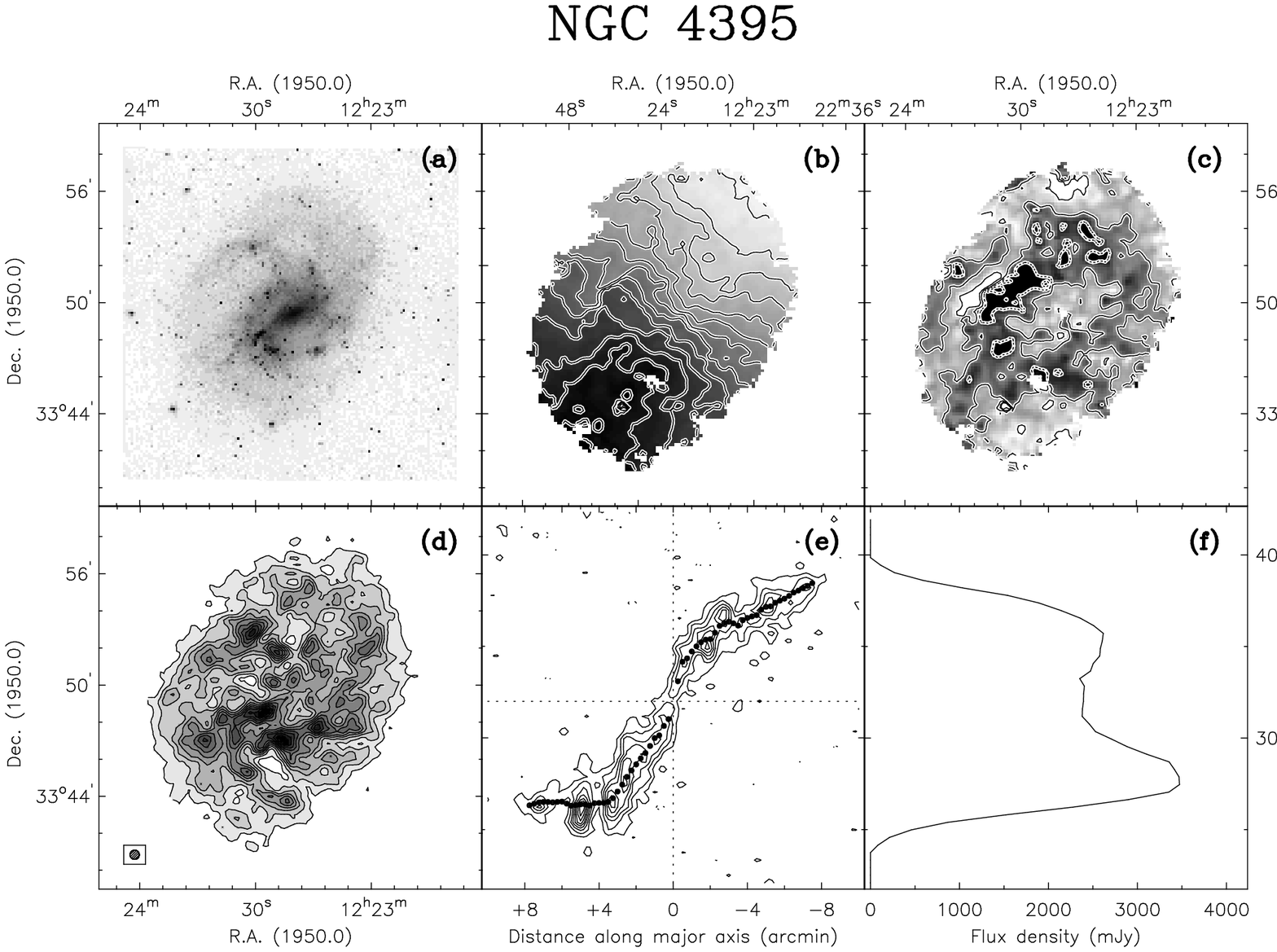}
\vskip0.7cm\null\vfill

\caption{{\bf Figure 1.} 
  Data panels for DDO~9 and NGC~4395.
  {\bf (a)} R-band image;
  {\bf (b)} Velocity field from Gauss fits, dark shading indicates approaching
     side, contour levels are 590 to 690 km s$^{-1}$ (DDO 9) and 260 to
     380 km s$^{-1}$ (NGC 4395), in steps of 10 km s$^{-1}$;
  {\bf (c)} Residual velocity field, dark shading indicates negative velocities,
     contours are spaced 7.5 km s$^{-1}$ apart, negative contours are
     dotted;
  {\bf (d)} Integrated HI map, the first contour level and the contour
     step are $2\cdot 10^{20}$ HI atoms cm$^{-2}$,
     the hatched circle in the lower left shows the
     $30''$ beam;
  {\bf (e)} Position-velocity diagram along the major axis, contours
     levels are $-2\sigma, 2\sigma, 4\sigma$, in steps of $2\sigma$,
     $\sigma=3.9$ mJy/beam for DDO~9
     and $3.1$ mJy/beam for NGC 4395, negative
     contours are dotted, the filled circles show the derived rotation
     curve;
  {\bf (f)} Global line profile.
  }

\null\vskip-1cm
\endfigure

In the mean time also the shape and brightness of the stellar discs
have been investigated in a quantitative way.  Rix \& Zaritsky (1995)
and Zaritsky \& Rix (1997) found that about 30\% of face-on field
spirals show significant lopsidedness in the $I$ and $K'$ bands.
Furthermore, galaxies with stronger lopsidedness have a $B$-band
luminosity excess compared to the luminosity predicted by the
Tully-Fisher relation. They suggest that the lopsidedness in the
galaxies in their sample is the result of recent accretion.  Rix \&
Zaritsky (1995) derived a wind-up time of about 1 Gyr for their
sample, contrasting the estimated lifetime based on their observed
frequency of lopsidedness of about 3 Gyr.  They also pointed out
that the winding problem can be avoided if one assumes that the
lopsidedness is caused by a lopsided potential, to which the gas and
the stars merely respond.

Aperture synthesis surveys of the HI distribution and kinematics of
large numbers of spiral galaxies now provide abundant and sufficiently
detailed information to investigate the nature of the asymmetries
indicated by the global profiles. We have noticed that in a large
number, perhaps the majority of cases, it is the kinematics rather
than the HI distribution that determines the asymmetry in the global
profiles.  Galaxies with such asymmetric global profiles often have
rotation curves that rise more slowly on one side of the galaxy than
on the other. Correspondingly, the velocity field has isovelocity
contours less curved on the side with the less steep rotation curve
than on the other.
 
Here attention is drawn to the phenomenon of lopsided kinematics and
two objects are shown to illustrate it. The HI observations are
described and it is shown that a lopsided potential perturbation can
produce the lopsided velocity fields. Finally, the question of the
frequency of kinematically lopsided galaxies is addressed. It is worth
noting that the two galaxies discussed here, as well as most of those
identified as kinematically lopsided, are isolated systems and do not
show signs of strong tidal interaction, nor is there clear evidence
for ongoing accretion of satellite galaxies.

\section{Data and Analysis}

\beginfigure*{2}
\line{
\epsfxsize=0.95\hsize\epsfbox{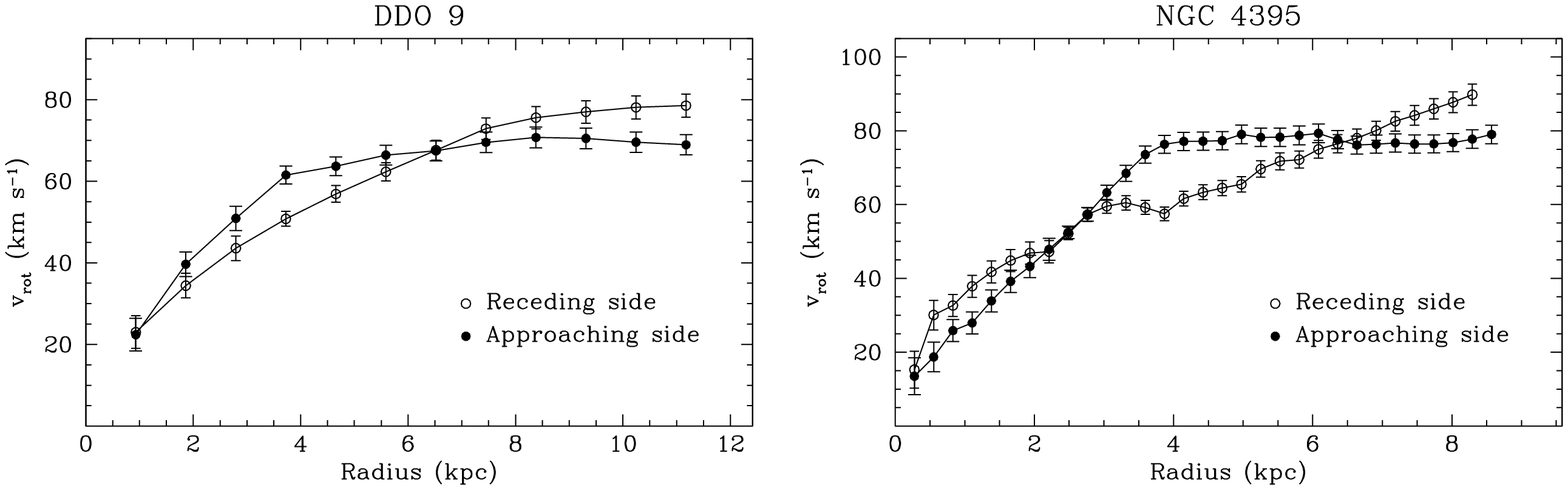}
}
\caption{{\bf Figure 2.} Rotation curves for the approaching and the
  receding sides separately. Open circles represent the receding side,
  filled circles the approaching side. The error bars are the formal
  errors in the fit, corrected for the beam size.}
\endfigure

Two objects, DDO~9 (UGC~731) and NGC~4395 (UGC~7524), have been
selected here as representative examples of kinematically lopsided
galaxies.  Fig. 1 shows an overview of the data.  The maps of the
integrated HI emission, the position-velocity diagrams along the major
axes and the global profiles originate from the Westerbork HI Survey
of Spiral and Irregular Galaxies (WHISP, see Kamphuis,
Sijbring \& van Albada 1996), the optical $R$-band images are from
Swaters \& Balcells (1998).

\begintable{1}
\caption{{\bf Table 1. Properties of DDO~9 and NGC~4395.} }
\halign{#\hfil & \quad\hfil#\quad & \hfil\quad#\cr
 & DDO 9 & NGC 4395 \cr
\noalign{\vskip2pt\hrule\vskip2pt}
Type & Im & Sm \cr
Centre: $\alpha(1950)$ & $1^h 7^m 46.7^s$ & $12^h 23^m 19.9^s$ \cr
\phantom{Centre: }$\delta(1950)$ & $49^\circ 20' 7''$ & $33^\circ 49' 26''$ \cr
Systemic velocity (km/s)& 638 & 320 \cr
Assumed distance (Mpc)& 12.8 & 3.8 \cr
Absolute magnitude $M_R$ & $-16.9$ & $-18.1$ \cr
R$_{25}$ (major $\times$ minor, arcmin) & $3.6\times 1.7$ &
 $14.4\times 9.1$ \cr
Inclination (degrees)& 57 & 46 \cr
Position angle (degrees)& 257 & 324 \cr
HI mass ($10^8\hbox{ M}_\odot$)& 7.4 & 9.1 \cr
}
\endtable

The radial velocity fields shown in Fig.\ 1, panels {\it b}, were
obtained from Gaussian curves fitted to the profiles.  The rotation
curve was derived by fitting tilted rings to the velocity field, using
the method described by Begeman (1989).  First, the kinematic centre
was determined for all rings from a fit with all parameters free. In
the case of NGC~4395, the average kinematic centre corresponded
closely to the optical centre, and because of its higher precision, we
used the latter. In the case of DDO~9, where the optical centre is not
well determined, we used the centre that minimized the residual
velocities, as is done in Schoenmakers, Franx \& de Zeeuw (1997;
hereafter SFdZ). By choosing the centre in this way, the difference
between the rotation curves of the approaching and receding sides is
minimized as well.  SFdZ show that it is important to keep the centre
fixed in these tilted ring fits, since a free centre will drift in
such a way as to make a real $m=2$ term disappear.

Another way of choosing the centre would have been to fix it in such a
way that the outer parts of the rotation curves of the receding and
approaching side coincide. This would mean that the gas in the outer
parts moves on circular orbits and that all the asymmetry is
concentrated in the inner parts. This would be a rather ad hoc choice,
especially for NGC~4395. It demonstrates, however, that in a
kinematically lopsided galaxy the centre cannot be determined uniquely
from its kinematics alone.

Next, the systemic velocity was determined from a tilted ring fit with
the centre fixed. The inclination and the position angle did not show
systematic trends with radius, and hence they were fixed to their
average values.  Finally, with the parameters of each ring fixed as
described above, the rotation curve was fitted with uniform spatial
weighting to the velocity field as a whole, giving the average
rotation curve.  Rotation curves were also derived separately for the
approaching and the receding sides in the same way.  The latter are
overplotted on the position-velocity diagrams in Fig. 1, panels {\it
  e}.  A model velocity field was constructed from the average
rotation curve, and subsequently subtracted from the observed one to
give the residual velocity field shown in Fig. 1, panels {\it c}.

Both galaxies are lopsided in their kinematics.  As can be seen in
Fig. 2, the rotation curves for the approaching and the receding sides
are distinctly different.  In both galaxies, the rotation curve of the
approaching side rises and subsequently flattens, whereas for the
receding side it continues to rise.  Obviously, the same pattern is
seen in the position-velocity diagrams (Fig. 1, panels {\it e}).
Accordingly, in the velocity fields the isovelocity contours are
curved more strongly on the approaching side where the rotation curve
becomes flat. This is best seen in NGC 4395.  The residual velocity
fields show clear systematic structure as well. In the case of DDO~9 a
twofold structure is visible; NGC~4395 shows a more or less circular
symmetric radial variation of the residual velocity, i.e., a radial
variation of the systemic velocity.  Note that the large-scale
distributions of the HI and the stars are fairly symmetric, despite the
lopsidedness in the kinematics.

To assist in the interpretation of the non-axisymmetric features of
the motion of the gas, the velocity fields have been decomposed into
harmonic components along individual rings, as found by the tilted
ring fit, following the approach of SFdZ. As detailed results for
several galaxies, including DDO~9 and NGC~4395, will be presented in a
forthcoming paper (Schoenmakers \& Swaters 1998), we restrict
ourselves here to the main points.  For both galaxies the dominant
terms are $m=0$ (circular symmetry), and $m=2$ (bi-symmetry). DDO~9
has strong $m=2$ terms, as expected from the twofold structure seen in
the residual velocity field. The $m=0$ term is weak. NGC~4395, on the
other hand, shows a strong radial variation of the $m=0$ term, and
only weak $m=2$ terms.

\section{Discussion}

\subsection{Kinematic lopsidedness versus lopsided potential}

In Section 2 we showed that kinematically lopsided galaxies have
velocity fields containing $m=0$ and $m=2$ terms.  In the epicycle
approximation, these harmonic terms in the velocity field can be
related to potential perturbations. SFdZ showed that the line-of-sight
velocity field contains $m-1$ and $m+1$ terms if the potential
contains a perturbation of harmonic number $m$.  Therefore, in the case
of an $m=1$ term in the potential (causing morphological
lopsidedness), the line-of-sight velocity field will contain an $m=0$
term and an $m=2$ term, not necessarily with the same amplitudes.
Hence, either or both of these terms should be visible in the residual
velocity field. The relative strengths of these terms depend on the
viewing angle of the perturbation $\phi_1$, as discussed below.

Expressions for the amplitudes of these harmonic components caused by
an $m=1$ perturbation in the potential are given by SFdZ in their
equation A28. We have used the SFdZ relations to create a model
velocity field.  Using the axisymmetric part of the potential as
derived from the mean rotation curve and a small, non-rotating ($\sim
10\%$) $m=1$ perturbation, the corresponding line-of-sight velocity
fields were calculated with different viewing angles.  Fig. 3 shows
the resulting velocity fields, projected with inclination
$i=55^\circ$, and viewing angle ranging from $\phi_1=0^\circ$ to
$\phi_1=90^\circ$, in steps of $30^\circ$. A viewing angle of
$\phi_1=0^\circ$ measured in the plane of the galaxy corresponds to a
perturbation whose major axis is aligned with the observed minor axis
of the system.  For a viewing angle $\phi_1=90^\circ$, the velocity
field shows the same characteristic asymmetry as seen in the observed
velocity fields in Fig. 1. As the viewing angle decreases, the
asymmetry becomes less pronounced.  At the same time, the residuals
change from being dominated by $m=0$ terms for $\phi_1=90^\circ$ to
$m=2$ terms for $\phi_1=0^\circ$.  For a viewing angle
$\phi_1=0^\circ$, the rotation curves derived from the approaching and
receding sides are identical, but the asymmetry between the
approaching and the receding sides is replaced by a more subtle one
between the near and the far side.  Such an asymmetry would be
difficult to find in observed velocity fields without a detailed
harmonic analysis.

The dynamical basis of the models used to construct the velocity
fields in Fig.\ 3 is weak. Therefore, no attempt has been made to
reproduce the velocity fields of DDO~9 and NGC~4395 quantitatively.
The models do illustrate, however, that a lopsided potential is
capable of producing an asymmetric velocity field, and therefore an
asymmetric rotation curve. The models do agree well with the
observations qualitatively, i.e. they show strong curvature of the
isovelocity contours on one side, corresponding to a relatively flat
rotation curve, and nearly straight isovelocity contours on the other
side, indicating solid body rotation. Furthermore, the models are
useful in illustrating the effects of variations in the viewing angle.

The above analysis demonstrates that the cause of a lopsided velocity
field can in principle be traced to a lopsided potential. It can be
shown (Schoenmakers and Swaters 1998) that a 5-10\%
perturbation in the potential will give rise to an approximately
10-20\% difference between the rotation curves of the approaching and
receding sides.  Hence, only a mild perturbation of the potential is
needed to produce a significant lopsidedness in the velocity field.
Note that a small $m=1$ potential perturbation will create
morphological lopsidedness as well, also in the case of a
non-selfgravitating component, since the orbits are no longer circular
(e.g., Jog 1997).

\beginfigure*{3}
\epsfxsize=0.9\hsize
\epsfbox{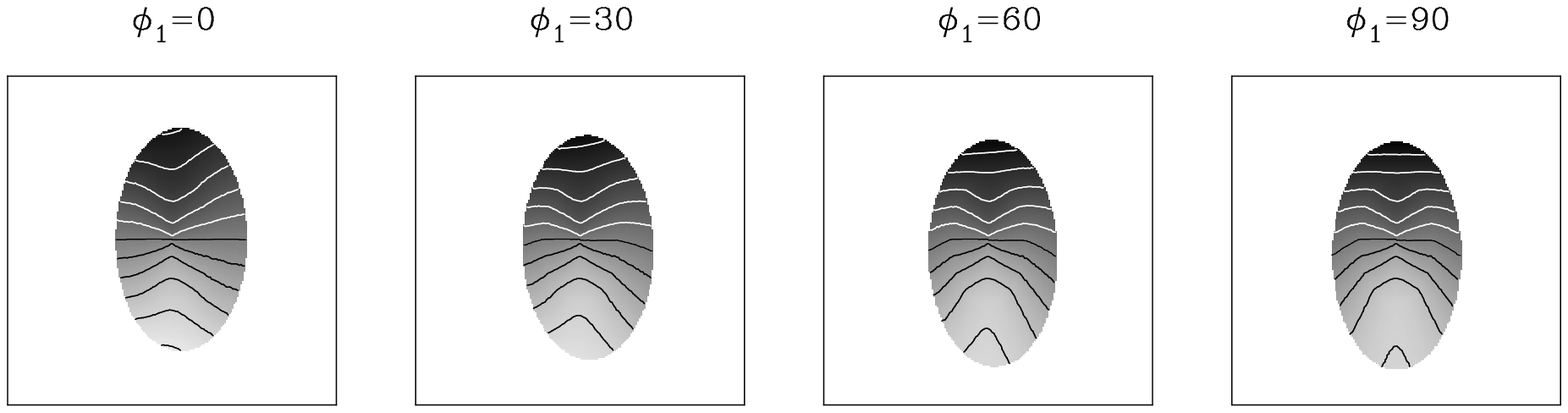}
\caption{{\bf Figure 3.} Model velocity fields derived from a lopsided
potential. All fields have an inclination of $55^\circ$ and a
viewing angle $\phi_1$ varying from $0^\circ$ to $90^\circ$ in steps
of $30^\circ$, as indicated above each panel. }
\endfigure

\subsection{Frequency of kinematically lopsided galaxies}

There are many other galaxies showing the same kinematical pattern as
the two objects described above. Striking examples are for instance
UGC~5459 and NGC~2770 (Rhee and van Albada, 1996).  The question now is:
how common is kinematic lopsidedness among disc galaxies?  

A first estimate of its frequency may be obtained from global profiles
(Richter \& Sancisi 1994, Haynes et al. 1998). This gives a
fraction of $~50\%$.  However, global profiles are influenced by both
the kinematics and HI distribution, as mentioned in the introduction.
The effect of kinematic lopsidedness on the global profile can be seen
clearly in Fig 1, panels {\it f}.  Usually, on the side where the
rotation curve becomes flat the corresponding horn in the global
profile has a higher peak and is narrower, whereas on the side of the
rising rotation curve it is wider and lower.  NGC 4395 is a good
example.  However, since both the density distribution and the
kinematics of the gas determine the shape of the global line profile,
the two may conspire in such a way as to hide the expected signature.
This is demonstrated in the case of DDO~9 (Fig. 1), where the flat
rotation curve side produces the lower intensity horn.  Furthermore,
if the lopsidedness is oriented along the minor axis, it will not show
up at all in the global profile.  Inspection of the global profiles
may, therefore, be helpful but is clearly insufficient for the purpose
of recognizing the presence of kinematic lopsidedness.

On the basis of published HI maps and major axis position-velocity
diagrams (Broeils \& van Woerden 1994, Rhee \& van Albada 1996), we
estimate the fraction of kinematically lopsided galaxies to be roughly
15\%. This fraction is a lower limit to the real fraction, because
position-velocity diagrams along the major axis will miss lopsidedness
directed along the minor axis. Similar estimates -at least 15\%, but
possibly up to 30\%- are also obtained from the recent HI survey
(Verheijen 1997) of a sample of about 40 spiral galaxies, which
despite their membership of the Ursa Major cluster are believed to be
representative for field galaxies.  It should be noted that in all
these selections by eye the fraction of lopsided galaxies will be
underestimated because mild asymmetries, asymmetries observed under
unfavourable viewing angles $\phi_1$, and asymmetries in nearly face-on
galaxies will be missed. Therefore, the fraction of kinematically
lopsided galaxies is probably at least 50\%.

Kinematic lopsidedness need not always be the dominant perturbation in
a galaxy.  Schoenmakers (1998) studied the harmonic analyses of nine
galaxies.  Seven of these were taken from Begeman (1987), who selected
large, regular, nearby spiral galaxies. The other two are LSB galaxies
(van der Hulst, Zwaan \& Bosma, priv.\ comm.).  Out of these nine
galaxies, seven show clear signs of kinematic lopsidedness in the
harmonic decomposition, but only in three of these seven the
lopsidedness is the dominant perturbation. In the other four the
lopsidedness would not have been detected without harmonic analysis.
It seems therefore that for a reliable estimate of the frequency and
distribution of amplitudes of kinematic lopsidedness it is necessary
to examine the harmonic decompositions with care.

\subsection{Origin of kinematic lopsidedness}

We have found from HI observations that the kinematic lopsidedness has
a well defined pattern that pervades the entire velocity field, and
that it is a common phenomenon among apparently isolated galaxies.
Optical (Rix and Zaritsky 1995, Zaritsky and Rix 1997) and infrared
(Block et al. 1994) photometry shows that also the stellar discs are
affected.  These facts suggest that lopsidedness is structural to the
disc, and that it is long-lived.  Using epicycle theory, kinematic
lopsidedness can be related to a lopsided potential.  In the case of
giant spirals the potential is probably governed by the disc, but in
late type systems such as DDO~9 and NGC~4395 the halo contribution may
well dominate. In such cases the lopsidedness may give information on
the structure of the halo and the location of the visible galaxy
inside it.

If the observed lopsidedness in galaxies is described as the result of
a stationary lopsided potential perturbation the winding problem
disappears, but the question arises of how to create and maintain such
a potential.  One possibility, of course, is a dynamical instability,
e.g., of the type proposed by Sellwood \& Merrit (1994). To create a
strong $m=1$ mode, their model requires strong counter-rotation.
Though some galaxies are known to have counter-rotating gas or stars
(e.g., Braun et al. 1994, Merrifield \& Kuijken 1994), most galaxies
do not (Kuijken, Fisher \& Merrifield 1996). Sellwood \& Valluri
(1997) suggest that the dark halo could possibly be the
counter-rotating component.

Lopsidedness may also be excited by interactions with a nearby
neighbour or by accretion, as suggested by Odewahn (1996) and Zaritsky
\& Rix (1997) on the basis of optical observations.  Since the
galaxies presented here, and a large fraction of the galaxies
inspected as described in section 3.2, do not appear to have
companions, this would require that lopsidedness excited in this way
must be a long lived phenomenon.  Simulations do not rule out this
possibility:  Walker, Mihos \& Hernquist (1996) have shown that
accretion of a small companion by a large disc galaxy can create large
scale asymmetries in the disc, that last up to about 1 Gyr.
Furthermore, Weinberg (1994) has shown that an $m=1$ distortion in a
King model is only weakly damped. In a halo-dominated galaxy, the
passage or accretion of another galaxy could excite such an $m=1$
perturbation in the halo that will persist for a long time (10-100
crossing times).

\section{Conclusions}

The presence of lopsidedness in the kinematics of disc galaxies has
been established and the pattern recognized. Two examples from recent
HI observations have been presented and discussed.  A rotation curve
which rises more steeply on one side of the galaxy than on the other
is a clear signature of kinematic lopsidedness. A large fraction of
spiral galaxies, probably at least 50\%, is kinematically lopsided.
Harmonic analyses of lopsided velocity fields show significant $m=0$
and $m=2$ terms.  Using epicycle theory, kinematic lopsidedness can be
related to a lopsided potential. A small $m=1$ perturbation in the
potential is sufficient to produce a significant lopsidedness in the
velocity field.

The present study has revealed that the lopsidedness in the kinematics
of spiral galaxies, as measured from the difference in rotation
velocities between the approaching and receding sides, may reach
amplitudes of order $10-20\%$. Larger kinematical disturbances have
not been found among the galaxies examined and may not exist, except
in case of strong tidal interactions. This seems to imply that
perturbations in the potential of isolated disc systems, as proposed
above as a mechanism to create kinematic lopsidedness, occur
frequently but that their amplitudes are unlikely to exceed about
$15\%$.

\section*{Acknowledgments}

The 21 cm line observations used in this paper are taken from the
WHISP survey (Westerbork HI survey of spiral and irregular galaxies)
carried out at the Kapteyn Astronomical Institute.  The WSRT is
operated by the Netherlands Foundation for Research in Astronomy with
financial support from the Netherlands Organization for Scientific
Research (NWO).

\end